\documentclass[review]{elsarticle}

\usepackage{lineno,hyperref}
\usepackage{underscore}
\usepackage{amssymb}
\usepackage{amsmath}
\usepackage{color}
\usepackage{epstopdf}
\usepackage{graphicx}
\usepackage{algorithm}  
\usepackage{algpseudocode}  

\journal{journal of computer science and technology}

\begin{document}

\begin{frontmatter}

\title{Identifying critical nodes in complex networks by graph representation learning}
\author{En-Yu Yu}
\address{Big Data Research Center, University of Electronic Science and Technology of China, Chengdu 611731, P. R. China}
\author{Yan Fu}
\address{Big Data Research Center, University of Electronic Science and Technology of China, Chengdu 611731, P. R. China}
\author{Duan-Bing Chen*}
\address{Big Data Research Center, University of Electronic Science and Technology of China, Chengdu 611731, P. R. China}
\address{The Research Base of Digital Culture and Media, Sichuan Provincial Key Research Base of Social Science, Chengdu 611731, P. R. China}
\address{Union Big Data Tech. Inc., Chengdu 610041, P. R. China}
\address{Correspondence should be addressed to dbchen@uestc.edu.cn}

\begin{abstract}
Because of its wide application background, critical nodes identification has become an important research content at the micro level of network science. Influence maximization is one of the main problems in critical nodes mining and is usually handled with heuristics. In this paper, a deep graph learning framework IMGNN is proposed and the corresponding training sample generation scheme is designed. The framework takes centralities of nodes in a network as input and the probability that nodes in the optimal initial propagation set as output. By training on a large number of small synthetic networks, IMGNN is more efficient than human-based heuristics in minimizing the size of
initial propagation nodes under the fixed infection scale. The experimental results on multiple real networks show that, compared with traditional non-iterative node ranking algorithms, IMGNN has the smallest proportion of initial propagation nodes under different infection probabilities when the final infection scale is determined. And the reordered version of IMGNN outperforms all the latest critical nodes mining algorithms.

\end{abstract}

\begin{keyword}
complex networks\sep critical nodes\sep graph learning \sep information maximization
\end{keyword}

\end{frontmatter}

\section{Introduction}
In our real life, many complex systems can be modeled as networks for analysis, such as power, road, and social networks\cite{Watts1998Collective, Ghosh2011Statistical, Breu2007The}. Therefore, understanding and controlling different networks is of great significance for social development. Complex networks are not only a form of data representation, but also a means of scientific research. At present, it has become a new hotspot in the field of network science to explain the different characteristics of networks at the micro level such as nodes and edges. Critical nodes identification is a valuable topic in network science and the results can also be applied in commodity marketing\cite{zhang2019discount}, opinion control\cite{masuda2015opinion} and malicious user detection\cite{wang2009understanding}. 

In the early stage of critical nodes mining in complex networks, many methods have been used to sort nodes in networks, such as degree centrality\cite{Bonacich1972Factoring}, Kshell\cite{kitsak2010identification}and PageRank\cite{Page1998The}. The significance of a single node can be effectively evaluated by these node ranking methods. However, with the advent of the era of big data, various networks in real life become increasingly complex, and the research focus gradually shifted from measuring the influence of a single node to maximizing the influence of a group of nodes\cite{li2017finding, bozorgi2017community, lv2019novel, liu2019identifying}. Influence maximization is a NP-hard problem\cite{Domingos2001Mining, Richardson2002Mining}. Therefore, researchers usually try to design some heuristic rules to obtain approximate results of optimal solutions by using the theory of graph and percolation. The simplest idea is to directly select the top-k nodes of node ranking algorithms. However, such approaches cannot reduce the overlap of influence between nodes\cite{Colizza2006Detecting}. Among all heuristic algorithms, the algorithms based on local attributes and iterative selection strategy perform very well. The so-called iterative selection means that a node is selected according to the existing network attributes and a certain strategy, then the network is updated and the next node is selected until a sufficient number of nodes are selected. In each iteration, VoteRank\cite{Zhang2016Identifying} select the most influential node according to the voting results of its neighbors, and then reduced the voting ability of its neighbors. As an improvement of VoteRank, NCVoteRank\cite{Kumar2020Identifying} select the most influential node by considering the coreness of its neighbors in the voting process. Similarly, EnRenew\cite{Guo2020Influential} selects the node with the maximum information entropy during each iteration. Wang et al.\cite{Wang2020Identifying} proposed an improved version of Kshell and selects the nodes with the maximum entropy from different shells. Although the above heuristic algorithms based on iterative selection can achieve a good balance between accuracy and cost, they are still based on some sort of heuristic designed by researchers, usually not applicable to all types of networks.

With the development of machine learning and the rise of deep learning, graph neural networks\cite{Wu2020A} (GNNs) emerged and gradually became a research hotspot in the field of graph computing\cite{Niepert2016Learning, Zhang2018Link, Ma2019Community}. Through graph neural networks, it can effectively learn the aggregation features of nodes in complex networks\cite{Hamilton2017Inductive, velickovic2018graph}. In this paper, to solve the problem of influence maximization, a deep graph learning framework IMGNN is proposed and the corresponding training sample generation scheme is designed. The framework takes centralities of nodes in a network as input and the probability that nodes in the optimal initial propagation set as output. The experimental results on one synthetic and five real networks show that, compared with PageRank, H-index, Degree and Kshell, IMGNN has the smallest proportion of initial propagation nodes under different infection probabilities when the final infection scale is greater than 80\%. In addition, Also, as a non-iterative selection algorithm, IMGNN can also be reordered by RINF algorithm\cite{yu2020re} to further improve its performance. And the reordered version of IMGNN outperforms VoteRank, EnRenew, Improved Kshell and NCVoteRank in most cases.

\section{Methods}
\subsection{SIR Spreading Model}
 For the problem of information maximization, SIR simulation model\cite{Lu2016Vital} is often used to measure the effectiveness of algorithms. In this model, nodes typically have three states: Susceptible, Infected and Recovered. In each iteration, susceptible nodes will be infected by the infected nodes in its neighbors with a probability of $\mu$. And then, infected nodes will go to recovered nodes with a probability $\beta$. This process will continue until all the infected nodes in the network become recovered. Generally speaking, when the importance of nodes is measured by SIR model, the recovery probability $\beta$ can be set to 1.

\subsection{Imformation Maximization with Graph Neural Networks (IMGNN)}
Aiming at problem of information maximization, an obvious task is to minimize the size of initial nodes when the propagation scale is fixed. For general complex networks, we cannot obtain the optimal solution by means of brute-force (the size of initial propagation nodes increases from small to large, and all possible combinations are exhausted). However, brute-force can be used to find the minimum set of initial propagation nodes whose final infection scale is greater than a certain threshold (e.g. 80\%) in a small network (there may be multiple optimal solutions). With the help of graph neural networks, by training in a large number of small synthetic networks and learning the features of nodes in the optimal propagation sets, we can quickly judge the probability of nodes belonging to the optimal propagation sets in large-scale networks, and then get critical nodes. Based on the above ideas, we put forward an effective deep learning framework IMGNN to minimize the number of initial propagation nodes when the final infection scale is fixed in static networks.The framework takes centralities of nodes in a network as input and the probability that nodes belong to the optimal propagation set(s) as output. The critical nodes can be obtained by selecting nodes in descending order of probability.

\begin{figure}[htp]
	\centering
	\includegraphics[width=14cm, height=9cm]{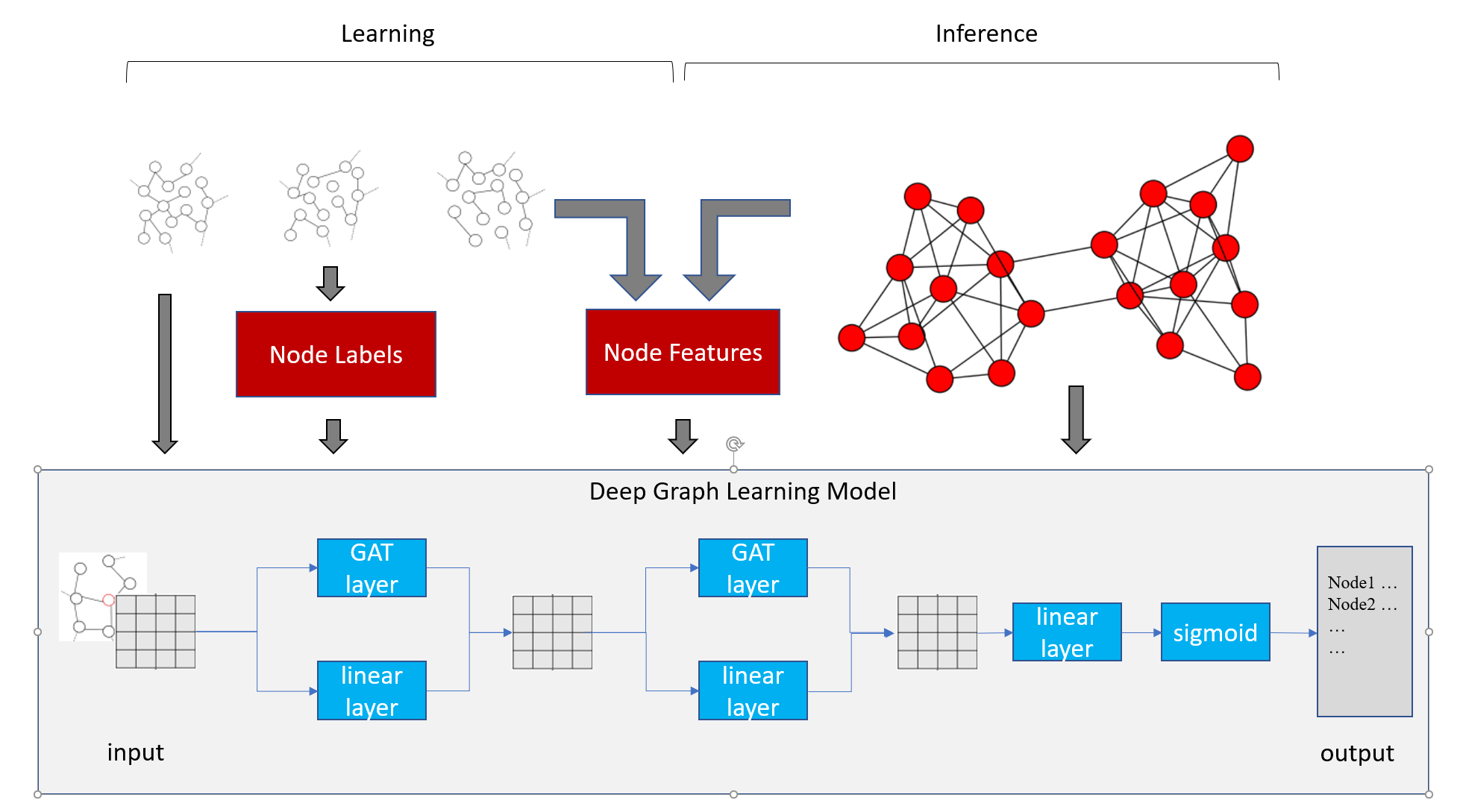}
	\caption{An illustration of IMGNN.}
	\label{fig:IMGNN}
\end{figure}

The main body of IMGNN is a deep graph learning model composed of the graph convolution layer and the linear layer. Its structure is shown in Figure \ref{fig:IMGNN}, which can be mainly divided into the following two modules:

\textbf{Training data generation}: IMGNN adopts supervised learning to train. Firstly, a large number of labeled training samples should be obtained. For this reason, we used BA model \cite{Barabasi1999Emergence} and ER model \cite{knuth1997Art} to generate 300 small networks with 15 nodes for model training. Among them, the networks generated by BA model include 50 networks with average degree of 4, 8 and 12 respectively, and the networks generated by ER model include 50 networks with random linking probability of 0.2 and 0.4 respectively. For each node in the network, some centralities and topological characteristics are calculated as the features of nodes. Specifically, we use the degree centrality and its $\chi^2$ value over the neighborhood, clustering coefficient and its $\chi^2$ value over the neighborhood, the PageRank value and the coreness as
input features. Chi-square value $\chi^2=\frac{(o-e)^2}{e}$, where $o$ represents the observed value (degree or clustering coefficient of the node), $e$ represents the expected value (the average degree or clustering coefficient of neighbors). The purpose of adopting the degree centrality (clustering coefficient) and its $\chi^2$ value over the neighborhood as features here is mainly to weaken the score of nodes which are directly connected to the nodes with high scores, and then reduce the probability of the nodes with the similar influence set being selected at the same time.
For example, the degree of node $A$ is 80 and the average degree of its neighbors is 100, the $\chi^2$ value of its degree over the neighborhood is 4.  the degree of node $B$ is 10 and the average degree of its neighbors is 1, the $\chi^2$ value of its degree over the neighborhood is 81. Although the degree of node $A$ is larger than that of node $B$, it can be found that node $B$ is more important in its neighborhood through the chi-square value transformation. Among the above six features, the degree centrality and its $\chi^2$ value over the neighborhood belong to first-order, the clustering coefficient and its $\chi^2$ value over the neighborhood belong to the second-order, the PageRank value and the coreness are global features. The combination of features with different granularity can describe nodes more effectively.

\begin{algorithm}[htb]  
	\caption{Search all optimal solutions.}  
	\label{alg:train}  
	\begin{algorithmic}[1]  
		\Require
		Graph $G = (V, E)$
		\Ensure
		Set of optimal solutions $S$\\
		\textbf{initialize}: $S= NULL$; $r=0$
		\While{$S == NULL$}
			\State $r= r + 1$
			\State $Cs$ is the set of all possible combinations of nodes (A total of $C_{|V|}^{r}$)
			\For{each combination $c$ in $Cs$}
				\If{the final infection scale is greater than 80\%}
				\State add $c$ to $S$
				\EndIf 
			\EndFor
		\EndWhile
	\end{algorithmic}  
\end{algorithm}  

In addition to the features of nodes, for the training data, we also need to obtain the label of each node. In this paper, nodes are assigned a numeric label which depends on their presence in the optimal propagation set(s). Since all training networks are micro-networks with 15 nodes, we can obtain all optimal solutions by the exhaustive method and the detailed process is shown in Algorithm \ref{alg:train}. Specifically, the label of each node is calculated as the number of optimal propagation sets it belongs to, divided by the total number of optimal propagation sets. If there is only one optimal set, we mark all nodes in the set as 1, and the other nodes in the network as 0. If there are two optimal set, we mark the nodes that belong to both sets as 1, the nodes that belong to a single set as 0.5, and the other nodes in the network as 0. In general, IMGNN considers the nodes that belong to most of the optimal propagation sets to be more critical than other nodes, and thus generates a large number of training data. Using these samples, the deep graph learning model can be effectively trained to obtain an appropriate way to aggregate node features.

\textbf{Deep graph learning model}: When considering the importance of nodes in the network, not only the attributes of nodes themselves, but also the influence of other nodes on them should be considered. Therefore, IMGNN uses the graph attention network (GAT) layer and the linear layer to learn the aggregation features and the own features of nodes respectively, and then superimpose them. As shown in Figure \ref{fig:IMGNN}, the graph convolution layer used by IMGNN is composed of the GAT layer and the linear layer. The input of the second convolution layer is the sum of the outputs of the GAT layer and the linear layer. After two continuous graph convolution layers, a linear fully connected layer is used to change the output dimension to 1. Finally, the output value is limited to $[0, 1]$ through a sigmoid active function. The graph attention network\cite{velickovic2018graph} introduced the attention mechanism\cite{jaderberg2015spatial}, the model first learned the weight adaptively for different neighbors, and then averaged it according to their weight. The transformation formula of node features in single-layer GAT network is as follows:

\begin{equation}
h^{'}_i(K)=\prod_{k=1}^{K} \sigma(\sum_{j\in \Gamma_i}\alpha_{ij}^{k}W^{k}h_j),
\end{equation}

\begin{equation}
\alpha_{ij}=\frac{e^{LeakyReLU(a([Wh_i][Wh_j]))}}{\sum_{k\in \Gamma_i}e^{LeakyReLU(a([Wh_i][Wh_k]))}},
\end{equation}

where $h_i$ and $h^{'}_i$ represent the features of node $i$ before and after the transformation respectively. $\Gamma_i$ is the first-order neighbors of node $i$. $K$ represents the amount of multi-head attentions. The parameters of the deep graph learning model were determined by grid search. In this paper, The number of output channels of GAT in the first and second layers is 10 and 20, respectively. Multiple attention parameters are 5 and 10, and each layer is coupled to a linear layer with the same number of input and output channels. The active function is ELU and the loss function is the squared loss function.

\section{Experiments}
In this paper, IMGNN trained 10 epoches on each training dataset with a learning rate of 0.001. All experiments were run on a single 1600 MHz GPU with 8G memory.

\subsection{Parameters Analysis}
In order to obtain the training set of IMGNN, we need to obtain all the optimal propagation sets of synthetic networks through brute force. At this time, we need to consider the influence of training infection probability $\mu_t$ on the training set and final results.

\begin{figure}[htbp]
	\centering
	\includegraphics[width=8cm, height=6cm]{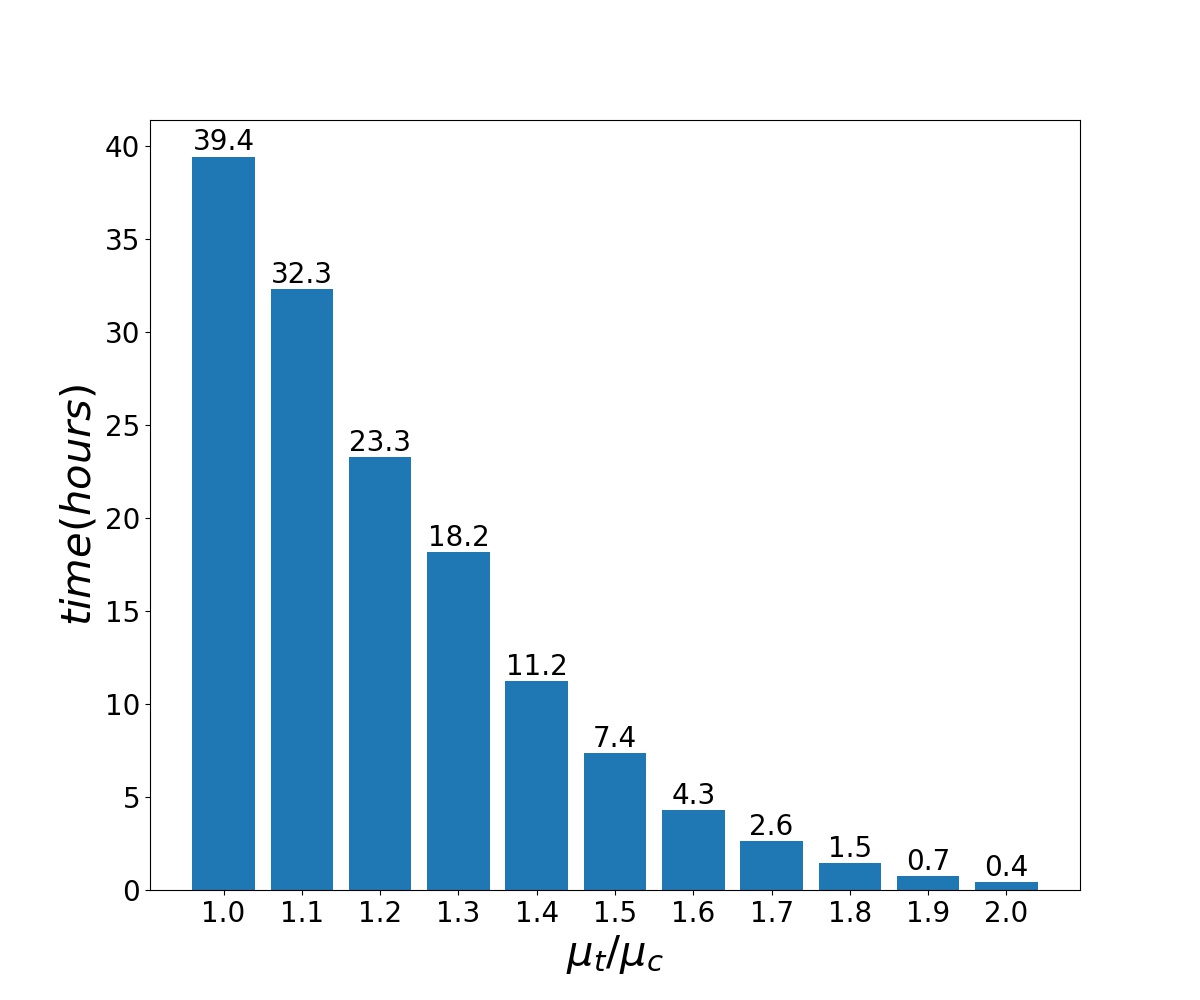}
	\caption{The time required to generate the training set varies with the training infection probability $\mu_t$, $\mu_c=\frac{\langle k \rangle}{\langle k^2 \rangle-\langle k \rangle}$.}
	\label{fig:RGATtrainTime}
\end{figure}

Since we use SIR model to simulate the transmission process of information in networks, the final infection scale needs to be obtained by means of 1000 simulation averaging. When the training infection probability $\mu_t$ is too small, the information is difficult to spread out, and the size of optimal propagation sets may exceed 50\%. Even in small networks, it will take a lot of time to generate node labels. As shown in figure \ref{fig:RGATtrainTime}, when the training infection probability $\mu_t/\mu_c =1$, it takes nearly 40 hours to generate node labels.

\begin{figure}[htbp]
	\centering
	\includegraphics[width=12cm, height=6cm]{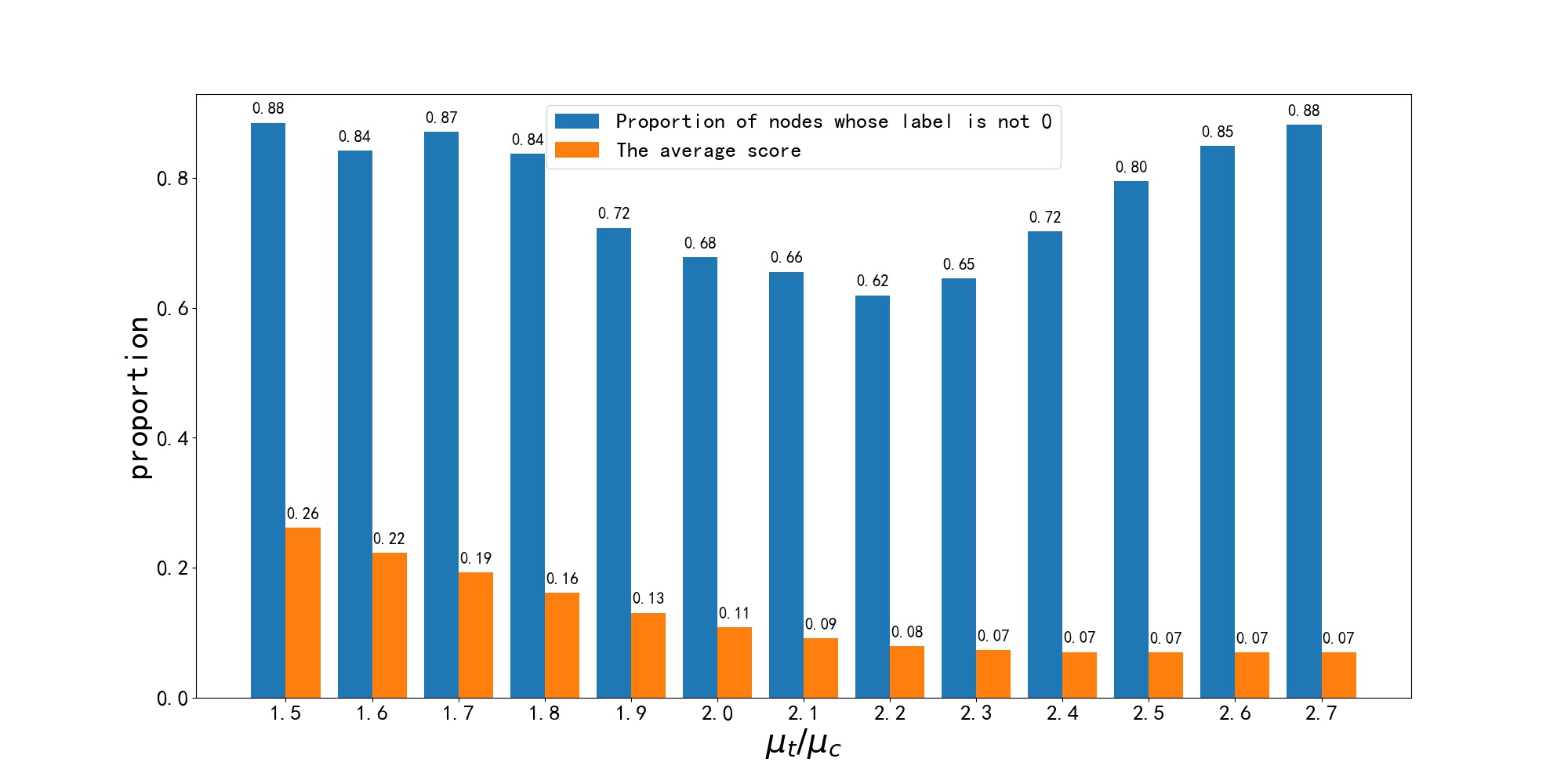}
	\caption{Effects of different training infection probabilities on the distribution of node labels, $\mu_c=\frac{\langle k \rangle}{\langle k^2 \rangle-\langle k \rangle}$.}
	\label{fig:RGATlabelNodes}
\end{figure}

When the training infection probability $\mu_t$ is too high, the information is easy to spread out. as shown in Figure \ref{fig:RGATlabelNodes}, when the training infection probability $\mu_t/\mu_c>2.2$, with the increase of $\mu_t$, the final infection scale of any node as the initial source is greater than 80\% and this results in that almost all optimal sets of training networks contain only 1 node. So the generated node labels are undifferentiated, with a large number of nodes are labeled $1/15$. When the training infection probability $\mu_t/\mu_c<2.2$, with the decrease of $\mu_t$, more nodes are needed as the initial source to make the final infection scale greater than 80\% and the size of the optimal sets will be larger and larger. This will also lead to higher and higher average scores for nodes.

In addition, it is not difficult to find that with the increase of training infection probability, the number of nodes with non-0 label presents a U-shaped distribution. When the training infection probability $\mu_t$ is small, optimal sets contains a large number of nodes. With the increase of $\mu_t$, the number of nodes required by optimal sets gradually decreases, which leads to the gradual decrease of the number of nodes with non-0 label. Until $\mu_t$ reaches a certain threshold
($\mu_t/\mu_c=2.2$), a single node can cause large-scale infection, while increasing $\mu_t$ will lead to more and more such nodes.

\begin{figure}[htbp]
	\centering
	\includegraphics[width=12cm, height=15cm]{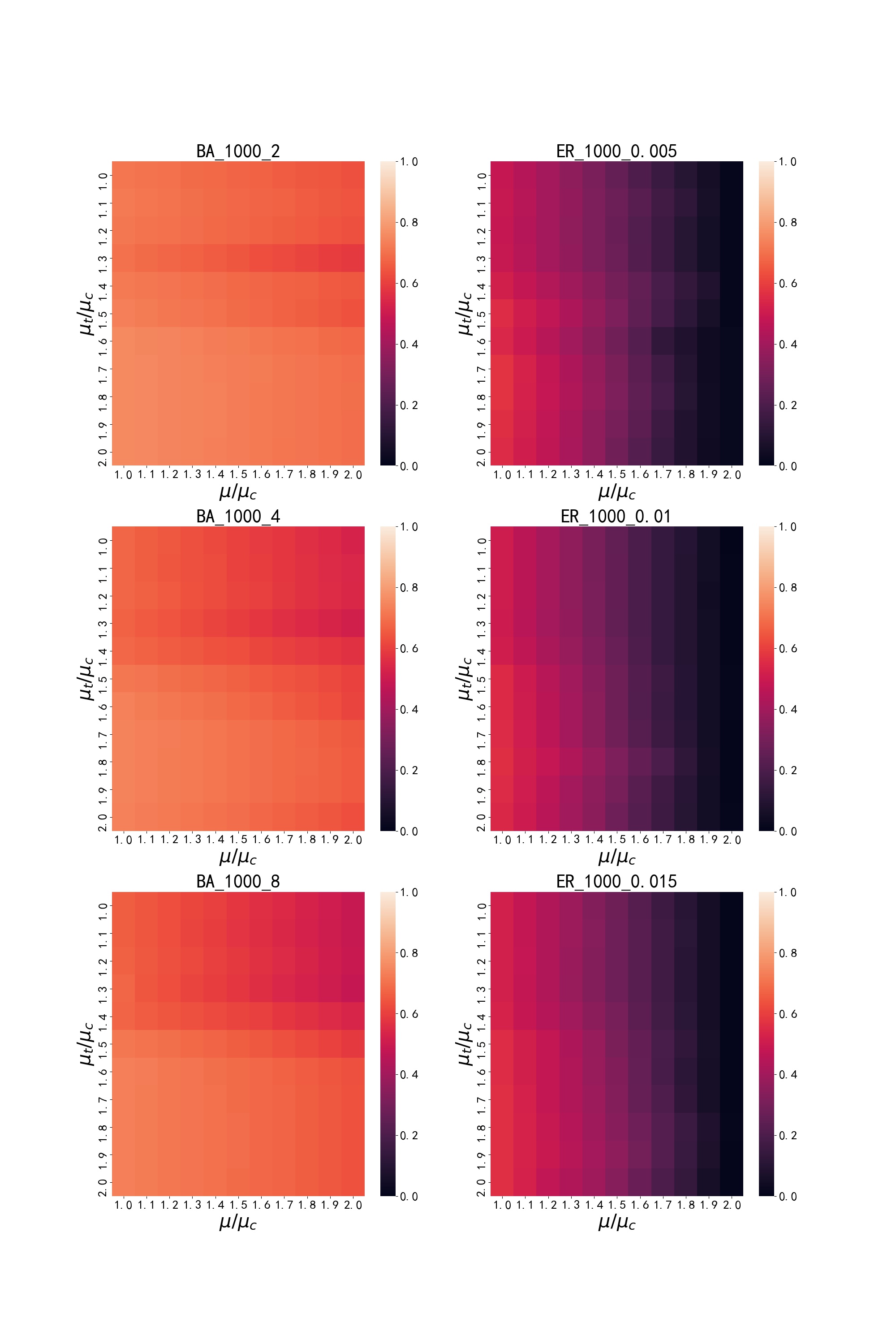}
	\caption{Effects of different training infection probabilities on the performance of IMGNN, $\mu_c=\frac{\langle k \rangle}{\langle k^2 \rangle-\langle k \rangle}$. BA\_1000\_2 is the network generated by BA model with 1000 nodes and the parameter of edge is 2. ER\_1000\_0.005 is the network generated by ER model with 1000 nodes and the parameter of edge is 0.005.}
	\label{fig:RGATheatmap}
\end{figure}

After analyzing the impact of the training infection probability $\mu_t$ on the time required to generate the training set and the distribution of node labels, It is not difficult to find that when $\mu_t/\mu_c \in [1, 2]$, the time cost of generating the training set and the data distribution are relatively reasonable. Within this range, the influence of training infection probability $\mu_t$ on the performance of IMGNN is shown in Figure \ref{fig:RGATheatmap}. In the experiment, six networks generated by the BA model and the ER model are used to test the final results of IMGNN. The ordinate represents the training infection probability when the training set is generated, and the abscissa represents
the real infection probability during the test. Each small square represents the minimum proportion of the initial propagation nodes when the final infection scale is greater than 80\%. Due to the high computational complexity of the SIR model, in the course of the experiment, we use a binary search algorithm to locate the minimum initial propagation node set. 

As shown in Figure \ref{fig:RGATheatmap}, it can be seen that as the real infection probability $\mu$ increases, the minimum proportion of the initial propagation nodes will continue to decrease. And the reduction of ER networks is more obvious (in BA networks, the propagation threshold $\mu_c=\frac{\langle k \rangle}{\langle k^2 \rangle-\langle k \rangle}$ is very small, resulting in insignificant changes). As the training infection probability $\mu_t$ decreases,  the minimum proportion of the initial propagation nodes will slightly decrease. Considering the uncertainty in the simulation process, it is almost negligible. In general, when $\mu_t/\mu_c \in [1, 2]$, the change of the training infection probability has little effect on the performance of IMGNN, and the results is better in the network generated by the ER model. 

\subsection{Datasets}
In order to effectively verify the performance of IMGNN in different networks, we will test it on six networks with different topology characteristics (the detailed values are shown in table \ref{data.IMGNN}). (1) BA: a scaling free network generated by BA model \cite{Barabasi1999Emergence}. (2) Jazz\cite{gleiser2003community}: In the cooperative network between jazz musicians, nodes represent musicians and edges represent cooperative relationships between two musicians. (3) USAir\cite{konect}: The network of flights between airports in the United States in 2010, with nodes representing airports and edges representing routes between two airports. (4) PGP\cite{konect:boguna}: a user interaction network based on Pretty Good Privacy (PGP) algorithm. (5) Sex\cite{rocha2011simulated}: A network of sexual activity in which nodes are women (sellers of sex) and men (buyers of sex), with edges representing sexual activity between two nodes. (6) Router\cite{spring2004measuring}: a router network, nodes represent routers, and connected edges represent information interaction between two routers.

\begin{table}
	\centering
	\caption{The basic properties of one synthetic network and five real networks. including the number of nodes ($N$) and edges ($M$), the max degree $k_{max}$, the average clustering coefficient $\langle C \rangle$ and the degree heterogeneity $H=\frac{\langle k^2 \rangle}{\langle k \rangle^2}$.}\label{data.IMGNN}
	\begin{tabular}{l llll lll}
		\hline
		Networks&$n$&$m$&$\langle k\rangle $&$k_{max}$&$\langle c \rangle$&$H$ & Connected \\ \hline
		BA      & 1000      &  9900     &  19.8       & 176    &  0.0618  & 1.7859 & Yes \\
		Jazz     & 198      &  2742     &  27.7       & 100    &  0.6174  & 1.3951 & Yes\\
		USAir      & 1574      & 17215     & 21.9       & 314    &  0.5042  & 5.1303& Yes \\
		PGP      & 10680      &  24316     &  4.6    & 205    &  0.2659  & 4.1464 & No \\
		Sex      & 16730      &  39044     &  4.7    & 305    &  0  & 6.0119 & No\\
		Router      & 5022      &  6258     &  2.5    & 106    &  0.0115  & 5.5031&  Yes\\
		\hline
	\end{tabular}
\end{table}

\subsection{Benchmark Methods}

In this paper, four well-known node ranking methods H-index, Kshell, PageRank and Degree are selected as benchmark methods to evaluate the performance of  IMGNN and four excellent algorithms based on iterative selection such as Improved Kshell, VoteRank, NCVoteRank and EnRenew are used to compare with the reordered version of IMGNN. 

The degree centrality\cite{Bonacich1972Factoring}:
\begin{equation}
DC(i) = \frac{k_{i}}{N-1},
\end{equation}
where $k_{i}$ the degree of node $i$.

PageRank\cite{Page1998The}:
\begin{equation}
PR_i(t) = (1-c)\sum_{j \in V/\{i\}}^{}a_{ji}\frac{PR_j(t-1)}{k_j}+\frac{c}{N},
\end{equation}
where $a_{ji}$ is an element of the adjacency matrix of the network.

The Kshell\cite{kitsak2010identification} of node $i$ is defined as the size of the largest complete subgraph containing node $i$.

The H-index\cite{Lu2016H-index}:
\begin{equation}
h(i)=H(k_{j_1},k_{j_2},...,k_{j_{k_i}}),
\end{equation}
where $k_{j_1},k_{j_2},...,k_{j_{k_i}}$ are degrees of node $i$'s neighbors. The goal of  H-index is to find a maximum integer $x$ such that there must be at least $x$ neighbors with degrees $\ge x$

RINF\cite{yu2020re} is a critical nodes mining algorithm that can reorder node ranking algorithms such as Degree, k-shell, PageRank, H-index and IMGNN. With the help of a special iterative selection strategy, the overlap of influence among nodes selected by initial algorithms can be effectively weakens and the performance of node ranking algorithms can be improved.

VoteRank\cite{Zhang2016Identifying} and NCVoteRank\cite{Kumar2020Identifying} get the score of a node according to its neighbors. In each iteration selection process, nodes vote to their neighbors, and select the node with the highest score to join the candidate set and reduce the voting capacity of its neighbors (The initial voting capacity is 1). In VoteRank, The score of node is the sum of the voting ability of its neighbors. In NCVoteRank,  The score is calculated by the normalized neighborhood coreness.

In EnRenew\cite{Guo2020Influential}, authors use the concept of node's information entropy:
\begin{equation}
E_i=-\sum_{j\in \Gamma_(i)}^{}\frac{k_j}{\sum_{l\in \Gamma_(i)}^{}k_l} \cdot log \frac{k_j}{\sum_{l\in \Gamma_(i)}^{}k_l}.
\end{equation}
And for each step, add the node with highest information entropy to the candidate set and reduce the information entropy of its neighbors.

In Improved Kshell\cite{Wang2020Identifying}, during each iteration, firstly, select the node with the highest information entropy from the node set with the highest coreness, and then select the node with the highest information entropy from the node set with the second highest coreness, This process continues until coreness drops to 1.

\subsection{Results}
Through the parameter analysis of IMGNN, it is not difficult to find that the model can be effectively trained by using the training set generated when the training infection probability $\mu_t/\mu_c \in [1, 2]$. Therefore, In the course of the comparative experiment, IMGNN is trained with the training set of $\mu_t/\mu_c =1.5$, and then the trained model is compared with other algorithms.

\begin{figure}[htbp]
	\centering
	\includegraphics[width=10cm, height=15cm]{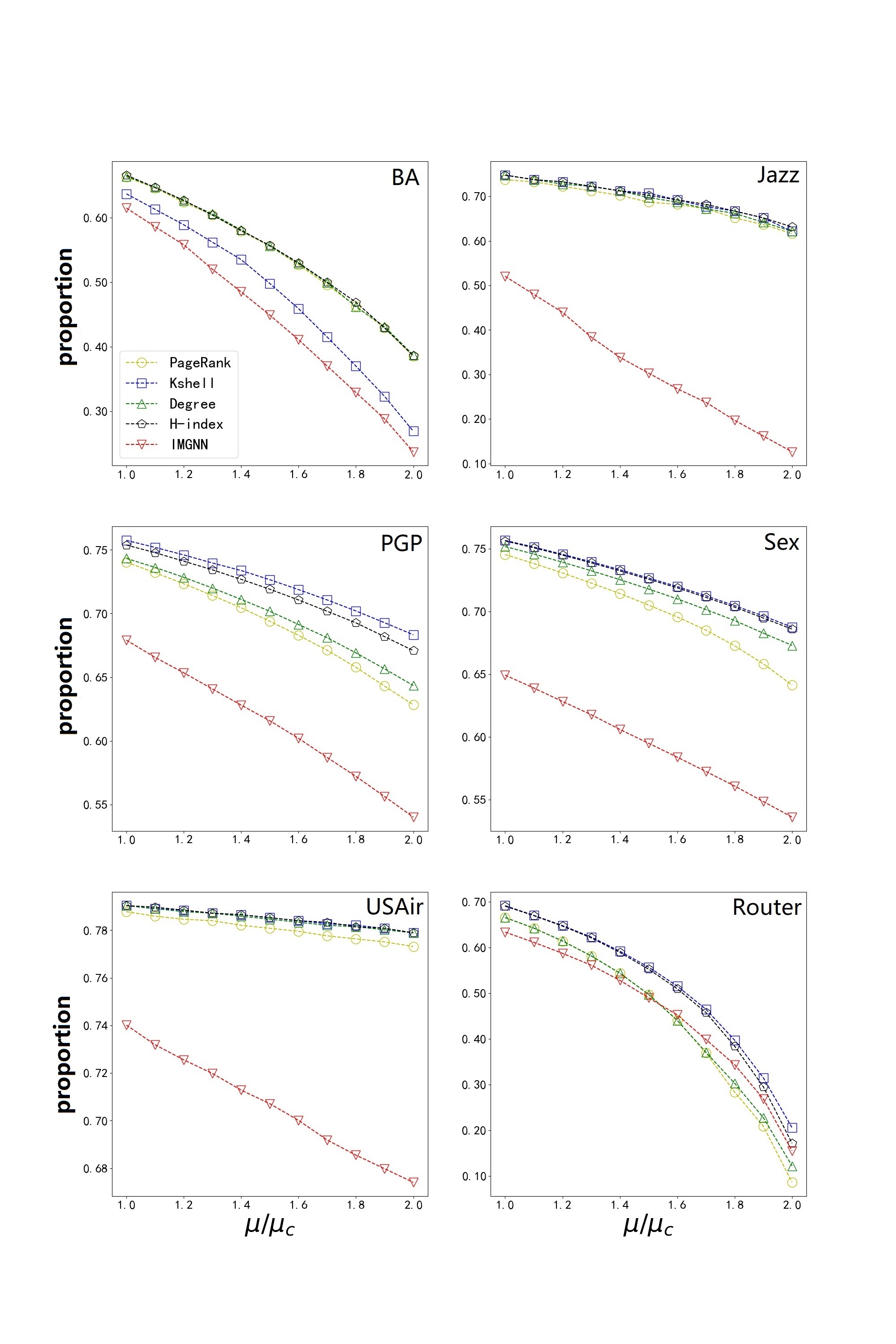}
	\caption{The proportion of the minimum initial propagation set varies with the infection probability $\mu$ when the final infection scale is greater than 80\% (node ranking algorithms), $\mu_c=\frac{1}{\langle k \rangle}$.}
	\label{fig:RGATmuO}
\end{figure}

First, we compare the proportion of the minimum initial propagation set required by the non-iterative algorithms when the final infection scale is greater than 80\% under different real infection probabilities. Due to the existence of unconnected networks in test networks, in order to reflect the difference of algorithms, we set the infection probability threshold $\mu_c=\frac{1}{\langle k \rangle}$. Figure \ref{fig:RGATmuO} shows the proportion of the minimum initial propagation set generated by non-iterative node ranking algorithms varies with the infection probability $\mu$.

As shown in Figure \ref{fig:RGATmuO}, IMGNN has the best performance on all networks except the Router, and in Router, IMGNN also outperforms other algorithms when the infection probability is low. In Jazz, the performance of IMGNN is much better than other algorithms, which is due to the obvious phenomenon of rich club and large average clustering coefficient. The nodes with higher ranking given by the traditional non-iterative node ranking algorithms are all in the large community, and the external nodes are ignored. The increase in the size of initial propagation set will not significantly increase the final infection scale. The same as Jazz is USAir, and USAir is an unconnected network, which leads to the fact that the critical nodes found by traditional algorithms are basically in the largest connected branch, and the proportion of the largest connected branch is less than 80\%. Therefore, even under a high infection probability, The proportion of the minimum initial propagation set is close to 80\% to reach the target. Experiments show that IMGNN can accurately judge the probability that the nodes in the network belong to optimal initial propagation set through learning on a large number of artificial networks.

\begin{figure}[htb]
	\centering
	\includegraphics[width=10cm, height=15cm]{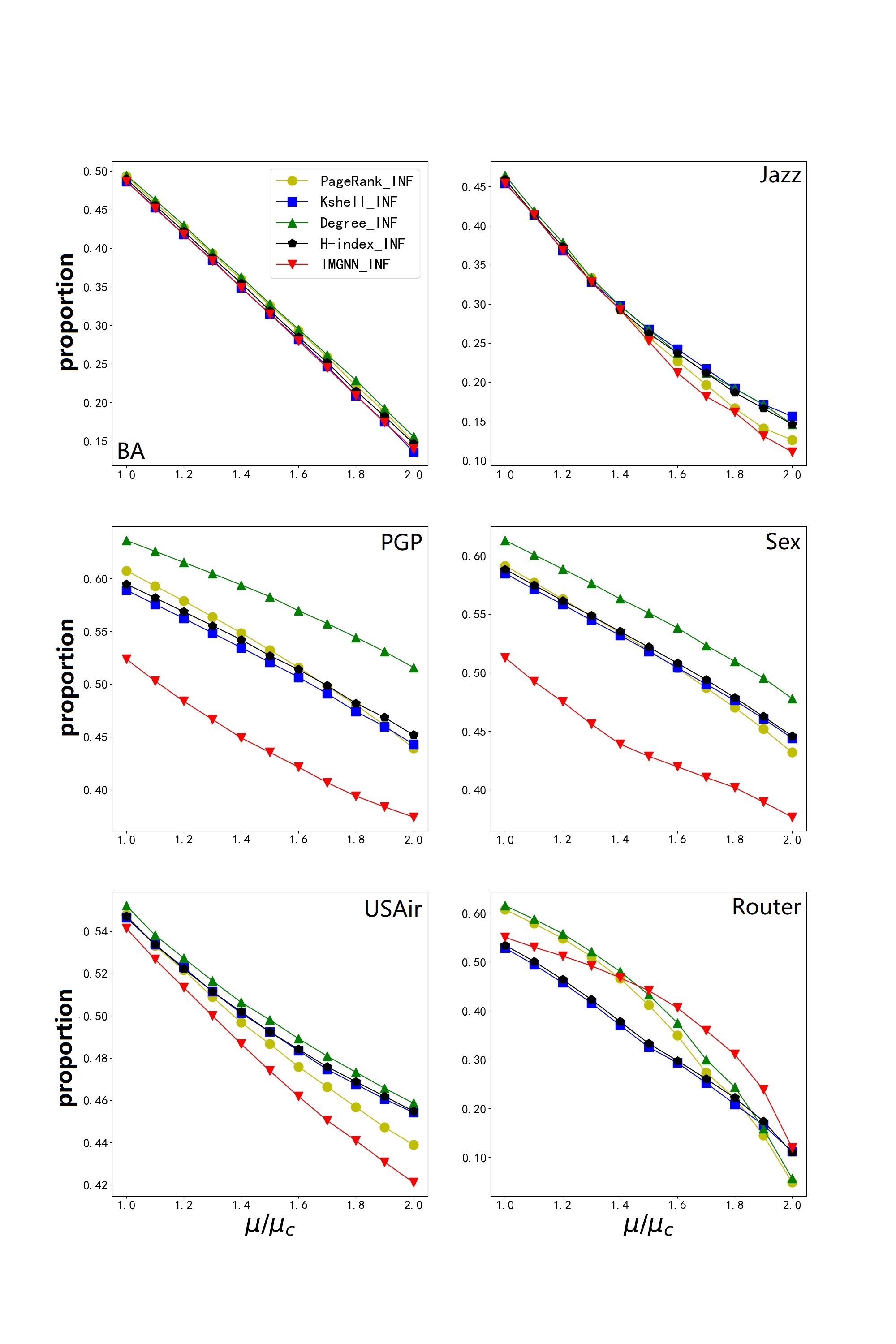}
	\caption{The proportion of the minimum initial propagation set varies with the infection probability $\mu$ when the final infection scale is greater than 80\% (reorder version of non-iterative node ranking algorithms),  $\mu_c=\frac{1}{\langle k \rangle}$.}
	\label{fig:RGATmuINF}
\end{figure}

Similar to the traditional non-iterative node ranking algorithms, IMGNN can give the score of each node (the probability of belonging to the optimal initial propagation set), which enables us to reorder it through RINF algorithm. Figure \ref{fig:RGATmuINF} shows the proportion of the minimum initial propagation set generated by reorder version of non-iterative node ranking algorithms varies with the infection probability $\mu$.

As shown in Figure \ref{fig:RGATmuINF}, After reordering by RINF, the performance of each algorithm is improved, and its average improvement is shown in Table \ref{tb.RGATCA}. Compared with other algorithms, IMGNN\_INF performs best on BA (no significant difference among different algorithms), Jazz, PGP, Sex and USAir. It can be seen from Table \ref{tb.RGATCA}, except for PGP, RINF has the smallest improvement on IMGNN, especially in Jazz network, which also reflects that IMGNN has weakened the scores of nodes which are directly connected to the nodes with high scores in the process of training. Compared with other algorithms, IMGNN has the least dependence on reordering and can directly achieve a good result. Experimental results show that RINF can improve the performance of IMGNN to some extent. And compared with other reordering algorithms, IMGNN\_INF still has advantages in most networks.

\begin{table}[htb]
	\centering
	\caption{Average improvement of node ranking algorithms under different infection probabilities. }\label{tb.RGATCA}
	\begin{tabular}{l llll ll}
		\hline
		Networks&IMGNN&Degree&PageRank&K-shell&H-index \\ \hline
		BA     & \textbf{0.1280}     &  0.2156     &  0.2185      & 0.1640    & 0.2257  \\
		Jazz      & \textbf{0.0514}      & 0.4100    & 0.4123      & 0.4146    & 0.4159  \\
		PGP     & 0.1725    &  \textbf{0.1186}     &  0.1611       & 0.2051    &  0.1905 \\
		Sex      &\textbf{0.1574}      &  0.1667    & 0.1846       & 0.2079    &  0.2037\\
		USAir      & \textbf{0.2291}      & 0.2839     & 0.2917    & 0.2898    &0.2894   \\
		Router      &\textbf{0.0541}     & 0.0613     &  0.0701   & 0.1865   & 0.1715  \\
		\hline
	\end{tabular}
\end{table}

Finally, we compare IMGNN\_INF with some latest critical nodes mining algorithms based on iterative selection strategy. Figure \ref{fig:RGATmuINF} shows the proportion of the minimum initial propagation set generated by critical nodes mining algorithms based on iterative selection strategy varies with the infection probability $\mu$.

\begin{figure}[htbp]
	\centering
	\includegraphics[width=10cm, height=15cm]{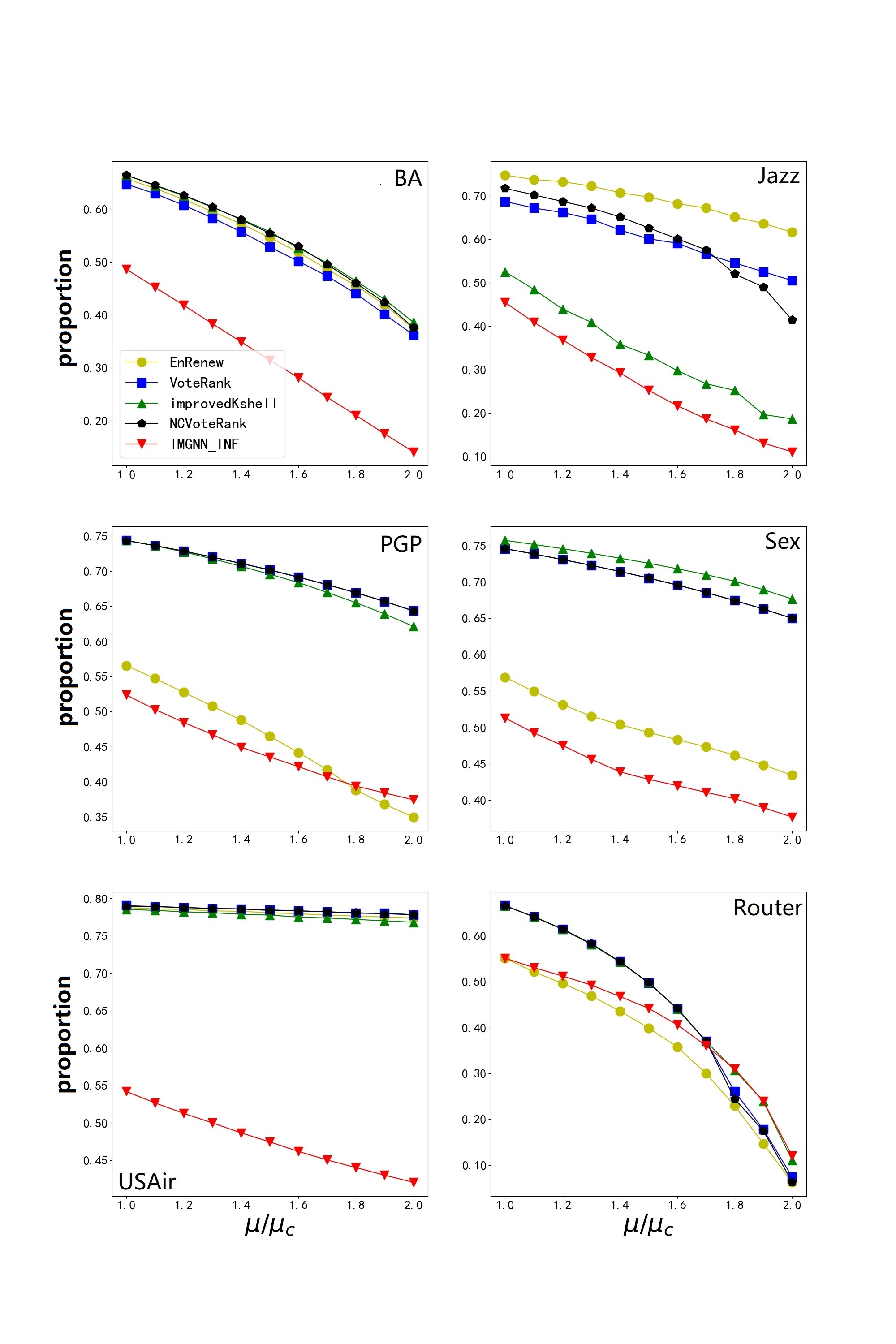}
	\caption{The proportion of the minimum initial propagation set varies with the infection probability $\mu$ when the final infection scale is greater than 80\% (critical nodes mining algorithms based on iterative selection strategy), $\mu_c=\frac{1}{\langle k \rangle}$.}
	\label{fig:RGATmuall}
\end{figure}

As shown in Figure \ref{fig:RGATmuall}, IMGNN\_INF performs better than other comparison algorithms on BA, Jazz, PGP (with most infection probabilities), Sex and USAir. The results on USAir network show that the existing critical nodes mining algorithms is still unable to deal with the network which is disconnected.

\section{Conclusions}
critical nodes mining in complex networks has very important application value in many fields such as public opinion control, network attack and defense, advertising and marketing. In this paper, aiming at the problem of minimizing the size of initial propagation nodes under the fixed infection scale, a deep graph learning framework IMGNN for critical nodes mining is proposed. Through training in a large number of small synthetic networks, IMGNN learns the characteristics of nodes belonging to the minimum initial propagation set, and can effectively evaluate the probability of nodes belonging to the set. The results of parameter analysis on synthetic networks show that IMGNN can be effectively trained by training set generated when the training infection probability $\mu_t/\mu_c \in [1, 2]$. Compared with traditional non-iterative node ranking algorithms, the experimental results show that IMGNN has the smallest proportion of initial propagation nodes under different infection probabilities when the final infection scale is greater than 80\%. In addition, the ranking results of IMGNN can also be reordered by RINF algorithm to further improve its performance, and the performance of the reordered results on most test networks exceeds that of all the latest critical nodes mining algorithms.

\clearpage

\section*{Acknowledgements}
This work is jointly supported by the National Natural Science Foundation of China under Grant Nos. 61673085, and by the Science Strength Promotion Programme of UESTC under Grant No. Y03111023901014006.

\bibliography{mybibfile}

\begin{thebibliography}{10}
\expandafter\ifx\csname url\endcsname\relax
  \def\url#1{\texttt{#1}}\fi
\expandafter\ifx\csname urlprefix\endcsname\relax\def\urlprefix{URL }\fi
\expandafter\ifx\csname href\endcsname\relax
  \def\href#1#2{#2} \def\path#1{#1}\fi

\bibitem{Watts1998Collective}
D.~J. Watts, S.~H. Strogatz, {Collective dynamics of 'small-world' networks},
  Nature 393~(6684) (1998) 440--442.
\newblock \href {http://dx.doi.org/10.1038/30918} {\path{doi:10.1038/30918}}.

\bibitem{Ghosh2011Statistical}
S.~Ghosh, A.~Banerjee, N.~Sharma, S.~Agarwal, N.~Ganguly, S.~Bhattacharya,
  A.~Mukherjee, {Statistical analysis of the Indian Railway Network: A complex
  network approach}, Acta Physica Polonica B, Proceedings Supplement 4~(2)
  (2011) 123--137.
\newblock \href {http://dx.doi.org/10.5506/APhysPolBSupp.4.123}
  {\path{doi:10.5506/APhysPolBSupp.4.123}}.

\bibitem{Breu2007The}
F.~Breu, S.~Guggenbichler, J.~Wollmann,
  \href{http://search.ebscohost.com/login.aspx?direct=true&db=ufh&AN=26313783&site=ehost-live&scope=site}{{The
  Benefits of Facebook “Friends:” Social Capital and College Students' Use
  of Online Social Network Sites}}, Journal of Computer-Mediated Communication
  12~(4) (2007) 1143--1168.
\newline\urlprefix\url{http://search.ebscohost.com/login.aspx?direct=true&db=ufh&AN=26313783&site=ehost-live&scope=site}

\bibitem{zhang2019discount}
T.~Zhang, P.~Li, L.-X. Yang, X.~Yang, Y.~Y. Tang, Y.~Wu, A discount strategy in
  word-of-mouth marketing, Communications in Nonlinear Science and Numerical
  Simulation 74 (2019) 167--179.

\bibitem{masuda2015opinion}
N.~Masuda, Opinion control in complex networks, New Journal of Physics 17~(3)
  (2015) 033031.

\bibitem{wang2009understanding}
P.~Wang, M.~C. Gonz{\'a}lez, C.~A. Hidalgo, A.-L. Barab{\'a}si, Understanding
  the spreading patterns of mobile phone viruses, Science 324~(5930) (2009)
  1071--1076.

\bibitem{Bonacich1972Factoring}
{Phillip Bonacich},
  \href{https://www.tandfonline.com/doi/abs/10.1080/0022250X.1972.9989806?journalCode=gmas20}{{Factoring
  and weighting approaches to status scores and clique identification}}, The
  Journal of Mathematical Sociology 2~(1) (1972) 113--120.
\newline\urlprefix\url{https://www.tandfonline.com/doi/abs/10.1080/0022250X.1972.9989806?journalCode=gmas20}

\bibitem{kitsak2010identification}
M.~Kitsak, L.~K. Gallos, S.~Havlin, F.~Liljeros, L.~Muchnik, H.~E. Stanley,
  H.~A. Makse, Identification of influential spreaders in complex networks,
  Nature physics 6~(11) (2010) 888--893.

\bibitem{Page1998The}
L.~Page, S.~Brin, {The anatomy of a large-scale hypertextual Web search
  engine}, Computer Networks 30~(1-7) (1998) 107--117.
\newblock \href {http://dx.doi.org/10.1016/s0169-7552(98)00110-x}
  {\path{doi:10.1016/s0169-7552(98)00110-x}}.

\bibitem{li2017finding}
R.-H. Li, L.~Qin, J.~X. Yu, R.~Mao, Finding influential communities in massive
  networks, The VLDB Journal 26~(6) (2017) 751--776.

\bibitem{bozorgi2017community}
A.~Bozorgi, S.~Samet, J.~Kwisthout, T.~Wareham, Community-based influence
  maximization in social networks under a competitive linear threshold model,
  Knowledge-Based Systems 134 (2017) 149--158.

\bibitem{lv2019novel}
Z.~Lv, N.~Zhao, F.~Xiong, N.~Chen, A novel measure of identifying influential
  nodes in complex networks, Physica A: Statistical Mechanics and its
  Applications 523 (2019) 488--497.

\bibitem{liu2019identifying}
D.~Liu, H.~Nie, J.~Zhao, Q.~Wang, Identifying influential spreaders in
  large-scale networks based on evidence theory, Neurocomputing 359 (2019)
  466--475.

\bibitem{Domingos2001Mining}
P.~Domingos, M.~Richardson, {Mining the network value of customers}, in:
  Proceedings of the Seventh ACM SIGKDD International Conference on Knowledge
  Discovery and Data Mining, 2001, pp. 57--66.
\newblock \href {http://dx.doi.org/10.1145/502512.502525}
  {\path{doi:10.1145/502512.502525}}.

\bibitem{Richardson2002Mining}
M.~Richardson, P.~Domingos, {Mining knowledge-sharing sites for viral
  marketing}, in: Proceedings of the ACM SIGKDD International Conference on
  Knowledge Discovery and Data Mining, 2002, pp. 61--70.
\newblock \href {http://dx.doi.org/10.1145/775047.775057}
  {\path{doi:10.1145/775047.775057}}.

\bibitem{Colizza2006Detecting}
V.~Colizza, A.~Flammini, M.~A. Serrano, A.~Vespignani, {Detecting rich-club
  ordering in complex networks}, Nature Physics 2~(2) (2006) 110--115.
\newblock \href {http://arxiv.org/abs/0602134} {\path{arXiv:0602134}}, \href
  {http://dx.doi.org/10.1038/nphys209} {\path{doi:10.1038/nphys209}}.

\bibitem{Zhang2016Identifying}
J.~X. Zhang, D.~B. Chen, Q.~Dong, Z.~D. Zhao, {Identifying a set of influential
  spreaders in complex networks}, Scientific Reports 6 (2016) 1--13.
\newblock \href {http://arxiv.org/abs/1602.00070} {\path{arXiv:1602.00070}},
  \href {http://dx.doi.org/10.1038/srep27823} {\path{doi:10.1038/srep27823}}.

\bibitem{Kumar2020Identifying}
S.~Kumar, B.~S. Panda, {Identifying influential nodes in Social Networks:
  Neighborhood Coreness based voting approach}, Physica A: Statistical
  Mechanics and its Applications 553 (2020) 124215.
\newblock \href {http://dx.doi.org/10.1016/j.physa.2020.124215}
  {\path{doi:10.1016/j.physa.2020.124215}}.

\bibitem{Guo2020Influential}
C.~Guo, L.~Yang, X.~Chen, D.~Chen, H.~Gao, J.~Ma, {Influential nodes
  identification in complex networks via information entropy}, Entropy 22~(2)
  (2020) 242.
\newblock \href {http://dx.doi.org/10.3390/e22020242}
  {\path{doi:10.3390/e22020242}}.

\bibitem{Wang2020Identifying}
M.~Wang, W.~Li, Y.~Guo, X.~Peng, Y.~Li, {Identifying influential spreaders in
  complex networks based on improved k-shell method}, Physica A: Statistical
  Mechanics and its Applications 554 (2020) 124229.
\newblock \href {http://dx.doi.org/10.1016/j.physa.2020.124229}
  {\path{doi:10.1016/j.physa.2020.124229}}.

\bibitem{Wu2020A}
Z.~Wu, S.~Pan, F.~Chen, G.~Long, C.~Zhang, P.~S. Yu, {A Comprehensive Survey on
  Graph Neural Networks}, IEEE Transactions on Neural Networks and Learning
  Systems 32~(1) (2021) 4--24.
\newblock \href {http://arxiv.org/abs/1901.00596} {\path{arXiv:1901.00596}},
  \href {http://dx.doi.org/10.1109/TNNLS.2020.2978386}
  {\path{doi:10.1109/TNNLS.2020.2978386}}.

\bibitem{Niepert2016Learning}
M.~Niepert, M.~Ahmad, K.~Kutzkov, {Learning convolutional neural networks for
  graphs}, in: 33rd International Conference on Machine Learning, ICML 2016,
  Vol.~4, 2016, pp. 2958--2967.
\newblock \href {http://arxiv.org/abs/1605.05273} {\path{arXiv:1605.05273}}.

\bibitem{Zhang2018Link}
M.~Zhang, Y.~Chen, {Link prediction based on graph neural networks}, Advances
  in Neural Information Processing Systems 2018-December (2018) 5165--5175.
\newblock \href {http://arxiv.org/abs/1802.09691} {\path{arXiv:1802.09691}}.

\bibitem{Ma2019Community}
X.~Ma, D.~Dong, Q.~Wang, {Community Detection in Multi-Layer Networks Using
  Joint Nonnegative Matrix Factorization}, IEEE Transactions on Knowledge and
  Data Engineering 31~(2) (2019) 273--286.
\newblock \href {http://dx.doi.org/10.1109/TKDE.2018.2832205}
  {\path{doi:10.1109/TKDE.2018.2832205}}.

\bibitem{Hamilton2017Inductive}
W.~L. Hamilton, R.~Ying, J.~Leskovec,
  \href{https://arxiv.org/abs/1706.02216}{{Inductive representation learning on
  large graphs}}, in: Advances in Neural Information Processing Systems, Vol.
  2017-Decem, 2017, pp. 1025--1035.
\newblock \href {http://arxiv.org/abs/1706.02216} {\path{arXiv:1706.02216}}.
\newline\urlprefix\url{https://arxiv.org/abs/1706.02216}

\bibitem{velickovic2018graph}
P.~{Veličković}, G.~{Cucurull}, A.~{Casanova}, A.~{Romero}, P.~{Liò},
  Y.~{Bengio}, Graph attention networks, in: International Conference on
  Learning Representations, 2018.

\bibitem{yu2020re}
E.~Yu, Y.~Fu, Q.~Tang, J.-Y. Zhao, D.-B. Chen, A re-ranking algorithm for
  identifying influential nodes in complex networks, IEEE Access 8 (2020)
  211281--211290.

\bibitem{Lu2016Vital}
L.~L{\"{u}}, D.~Chen, X.~L. Ren, Q.~M. Zhang, Y.~C. Zhang, T.~Zhou, {Vital
  nodes identification in complex networks}, Physics Reports 650 (2016) 1--63.
\newblock \href {http://arxiv.org/abs/1607.01134} {\path{arXiv:1607.01134}},
  \href {http://dx.doi.org/10.1016/j.physrep.2016.06.007}
  {\path{doi:10.1016/j.physrep.2016.06.007}}.

\bibitem{Barabasi1999Emergence}
A.~L. Barab{\'{a}}si, R.~Albert, {Emergence of scaling in random networks},
  Science 286~(5439) (1999) 509--512.
\newblock \href {http://arxiv.org/abs/9910332} {\path{arXiv:9910332}}, \href
  {http://dx.doi.org/10.1126/science.286.5439.509}
  {\path{doi:10.1126/science.286.5439.509}}.

\bibitem{knuth1997Art}
D.~E. Knuth, The art of computer programming, Vol.~3, Pearson Education, 1997.

\bibitem{jaderberg2015spatial}
M.~Jaderberg, K.~Simonyan, A.~Zisserman, et~al., Spatial transformer networks,
  Advances in neural information processing systems 28 (2015) 2017--2025.

\bibitem{gleiser2003community}
P.~M. Gleiser, L.~Danon, Community structure in jazz, Advances in complex
  systems 6~(04) (2003) 565--573.

\bibitem{konect}
J.~Kunegis, \href{http://konect.uni-koblenz.de/networks}{{{\{}KONECT{\}} --
  {\{}The{\}} {\{}Koblenz{\}} {\{}Network{\}} {\{}Collection{\}}}}, in: Proc.
  Int. Conf. on World Wide Web Companion, 2013, pp. 1343--1350.
\newline\urlprefix\url{http://konect.uni-koblenz.de/networks}

\bibitem{konect:boguna}
M.~Boguñá, R.~Pastor-Satorras, A.~Díaz-Guilera, A.~Arenas, Models of social
  networks based on social distance attachment, Phys. Rev. E 70~(5) (2004)
  056122.

\bibitem{rocha2011simulated}
L.~E. Rocha, F.~Liljeros, P.~Holme, Simulated epidemics in an empirical
  spatiotemporal network of 50,185 sexual contacts, PLoS computational biology
  7~(3) (2011) e1001109.

\bibitem{spring2004measuring}
N.~Spring, R.~Mahajan, D.~Wetherall, T.~Anderson, Measuring isp topologies with
  rocketfuel, IEEE/ACM Transactions on networking 12~(1) (2004) 2--16.

\bibitem{Lu2016H-index}
L.~L{\"{u}}, T.~Zhou, Q.~M. Zhang, H.~E. Stanley, {The H-index of a network
  node and its relation to degree and coreness}, Nature Communications 7 (2016)
  1--7.
\newblock \href {http://dx.doi.org/10.1038/ncomms10168}
  {\path{doi:10.1038/ncomms10168}}.

\end{thebibliography}

\end{document}